\documentclass[jap,preprint,showpacs,showkeys, preprintnumbers,amssymb]{revtex4}
\usepackage{graphicx}
\usepackage{amssymb}
\usepackage{amsmath}
\usepackage{bm}
\begin{document}
\setcounter{page}{0}
\title{Mangetic phase transition for three-dimensional Heisenberg weak random anisotropy model: Monte Carlo study}
\author{Ha M. \surname{Nguyen}}
\email{nmha@ess.nthu.edu.tw}
\author{Pai-Yi \surname{Hsiao}}
\email{pyhsiao@ess.nthu.edu.tw}
\affiliation{Department of Engineering and System Science, National Tsing Hua University, Hsinchu, Taiwan 30013, R.O.C}
\date{\today}

\newpage
\begin{abstract}
Magnetic phase transition (MPT) to magnetic quasi-long-range order
(QLRO) phase in a three-dimensional Heisenberg weak ($D/J=4$) random anisotropy (RA) model is investigated by Monte Carlo simulation. The isotropic and cubic distributions of RA axes are considered for simple-cubic-lattice systems.
Finite-size scaling analysis shows that the critical couplings for the former
and latter are $K_{\rm c}= 0.70435(2)$ and $K_{\rm c}=0.70998(4)$, respectively. While the critical exponent $1/\nu =1.40824(0)$ is the same for both cases. A second-order MPT to the QLRO phase is therefore evidenced to be possible in favor with the existence of the QLRO predicted by recent functional renormalization group theories.
\end{abstract}

% \pacs{73.43.Cd, 74.25.Ha, 75.30.Cr, 75.40.Gb, 75.50.Lk, 77.80.Bh, 78.20.Bh}

\keywords{ Magnetic phase transition, random anisotropy model, amorphous magnets, Monte Carlo simulation, finite-size scaling}

\maketitle
Rare-earth amorphous magnets are among magnetic glasses where the randomly-quenched disorder is in the form of intrasite random anisotropy \cite{ref1}. This characteristic was first realized by Harris, Plischke, and Zuckermann in the Hamiltonian of their random anisotropy model (RAM) \cite{ref2},
\begin{equation}
\mathcal{H}=-J\sum_{\langle i,j\rangle}{\vec{S}_{i}\cdot\vec{S}_{j}}-D\sum_{i}{(\hat{a}_{i}\cdot\vec{S}_{i})^{2}},
\label{eq1}
\end{equation}
where $\vec{S}_{i}$ is a unit vector representing the Heisenberg spin on site $i$ of a simple cubic lattice of size $L$, which is subject to a local onsite anisotropy of strength $D>0$ and uniaxially-easy axis represented by a randomly-quenched unit vector $\hat{a}_{i}$, and which can interacts with its six nearest neighbors $\vec{S}_{j}$ via ferromagnetic exchange interaction of strength $J>0$.  The degree of such a topological disorder is characterized by the disorder strength $D/J$ and the distribution of the easy axes $p(\hat{a}_{i})$ which is either of an isotropic form
\begin{equation}  
            p(\hat{a}_{i})=\frac{1}{\int{d^{3}\hat{a}_{i}}}=\frac{\Gamma(3/2)}{2\pi^{3/2}},
            \label{eq2}
        \end{equation}
where $\Gamma(x)$ is Euler gamma-function, or of an anisotropic (cubic) form
\begin{equation}  
            p(\hat{a}_{i})=\frac{1}{6}\sum_{\alpha=1}^{3}{[\delta^{(3)}(\hat{a}_{i}-\hat{e}_{\alpha})+\delta^{(3)}(\hat{a}_{i}+\hat{e}_{\alpha})]},
            \label{eq3}
        \end{equation}
where $\delta(y)$ is Dirac $\delta$-function and $\hat{e}_{\alpha}$ are Cartesian unit vectors. Vector $\hat{a}_{i}$ is equally likely to point in any direction of the space for the former case while it points with the same probability of the latter in one of six directions along three Cartesian coordinate axes.

RAM is the prototypical model system of a majority of amorphous magnets whose magnetic properties have been intensively investigated since around 1973. The most interesting questions about the model to which answers are still not completely clear or  are contradictory concern (i) the nature of magnetic phase ordering at low temperatures and (ii) the nature of magnetic phase transitions (MPTs). According to Imry-Ma argument \cite{ref3}, {\it long-range order} (LRO) is destroyed in systems with a continuous symmetry in $d<4$ spatial dimensions by even arbitrarily weak disorder of the form of either random field (RF) \cite{ref3} or random anisotropy (RA) \cite{Pelcovits_1978}. Amorphous magnetic ordering (AMO) may be intuitively described by one of the non-collinear spin structures (NCSSs) suggested by Coey \cite{1978-Coey-JAP}. The feature of the NCSSs is the possibility that spins within a domain may be frozen into more or less random orientations. Chudnovsky {\it et al.} \cite{Chudnovsky} proposed a phenomenological theory based on the Hamiltonian in Eq. (\ref{eq1}) for the description of AMO. Their theory shows that the characteristic of AMO depends crucially on the parameter $\Lambda=(D/J)(R_{\rm c}/a)^{2}$, where $R_{\rm c}$ is the scale of the spatial correlation of the easy axes and $a$ is the atomic spacing. Of special interest is the prediction of the {\it correlated spin glass} (CSG) phase for weak RA, i.e. $\Lambda \ll 1$. The CSG exhibits a smooth rotation of the magnetization over the volume so that the directions of the magnetization are ferromagnetically correlated on a quite large length $R_{\rm D}\gg R_{\rm c}$, (e.g., $R_{\rm D}\sim \Lambda^{-2}R_{\rm c}=(J/D)^{2}a^{4}/R_{\rm c}^{3}$ for $d=3$). Although the ferromagnetic ordering exists on the scale $R_{\rm D}$ there are no sharp boundaries between the ferromagnetic regions, i.e. no sharp domain walls. In zero field, the net magnetization of the CSG is zero and the susceptibility is very large but finite $\chi =\frac{1}{8}(\Lambda/4)^{-4} \sim (J/D)^{4}(a/R_{\rm c})^{8}$. For strong RA, when $\Lambda \gg 1$, a spin-glass-like (SG) state, the sperromagnet (SPM), is found with $R_{\rm D}\sim R_{\rm c}$. Recent theoretical work by Feldman \cite{Feldman2000} using the functional renormalization group (FRG) method has shown that in many cases LRO is prohibited in such the systems of continuous symmetry as RF and RA glasses, but instead {\it quasi-long range order} (QLRO) can emerge in the $d=4-\epsilon$ dimensions for the presence of weak disorder. The QLRO is peculiarly characterized by a power-law correlation function and the average value of the order parameter (e.g. the net magnetization) over the volume being zero in zero field. The QLRO is more common in impure systems and not prohibited by non-Abelian symmetry. It is worth noting that there seems to appear a disagreement between the CSG and the QLRO theories that
the CSG indicates large but finite correlation length and susceptibility while they are infinite for the QLRO. In our opinion, this paradox may be removed by the fact independently predicted in the theories by Cochrane {\it et al.} \cite{Cochrane} and by Els\"asser {\it et al.} \cite{Elsasser} that no correlation indeed exists between the easy axes of neighboring sites. One, therefore, should take the limit $R_{\rm c}\rightarrow 0$ for $R_{\rm D}$ and $\chi$ in the CSG theory above to approach the infinity.

QLRO was clearly evidenced in a number of numerical experiments using optimal Monte Carlo (MC) methods \cite{Fisch,Itakura2003}. In particular, Itakura \cite{Itakura2003} has recently investigated the 3D Heisenberg RAM in Eq. (\ref{eq1}) with the easy-axis distribution in Eq. (\ref{eq2}) and found that the spin-spin correlation function for the low temperature phase can be described by $G(r)\propto r^{-\eta-1}\exp(-r/\xi)$ with a nonuniversal exponent $\eta$. The inverse of the correlation length $\xi$ is finite for large values of $D/J$ and vanishes when $D/J\le 5$ so that a nonuniversal QLRO emerges with the correlation function of the frozen power-law form $G(r)\propto r^{-\eta-1}$. While a second-order MPT to the QLRO for a weak RA case $D/J=1$ of the 3D XY RAM was found by R\"o\ss{}ler using MC simulations \cite{Robler1999}, whether the crossover from the paramagnetic state to the QLRO, which was observed by Itakura \cite{Itakura2003} for a weak RA case $D/J=4$ of the 3D Heisenberg RAM, is a true second-order MPT is still an open question.

In this paper, we present our results of the MPT and critical behaviors for the case $D/J=4$ of the 3D Heisenberg RAM in Eq. (\ref{eq1}) similar to that of Itakura's work. Systems of the simple cubic lattice with the periodic boundary condition are investigated. We, however, consider both cases of the distribution of the easy axes as in Eqs. (\ref{eq2}) and (\ref{eq3}) which are called isotropic radom anisotropy model (IRAM) and anisotropic radom anisotropy model (ARAM), respectively. The reason for this consideration of the easy-axis distributions shall be cleared up in the sequel. Since we do not study the glassy phase transition but the MPT we, therefore, focus only on the magnetic order parameter, that is the net magnetization $m=\sqrt{M_{x}^{2}+M_{y}^{2}+M_{z}^{2}}$ where $M_{\alpha}=L^{-3}\sum_{i}S_{i\alpha}$.

\begin{figure}
       \begin{center}
         \resizebox{!}{!}{\includegraphics{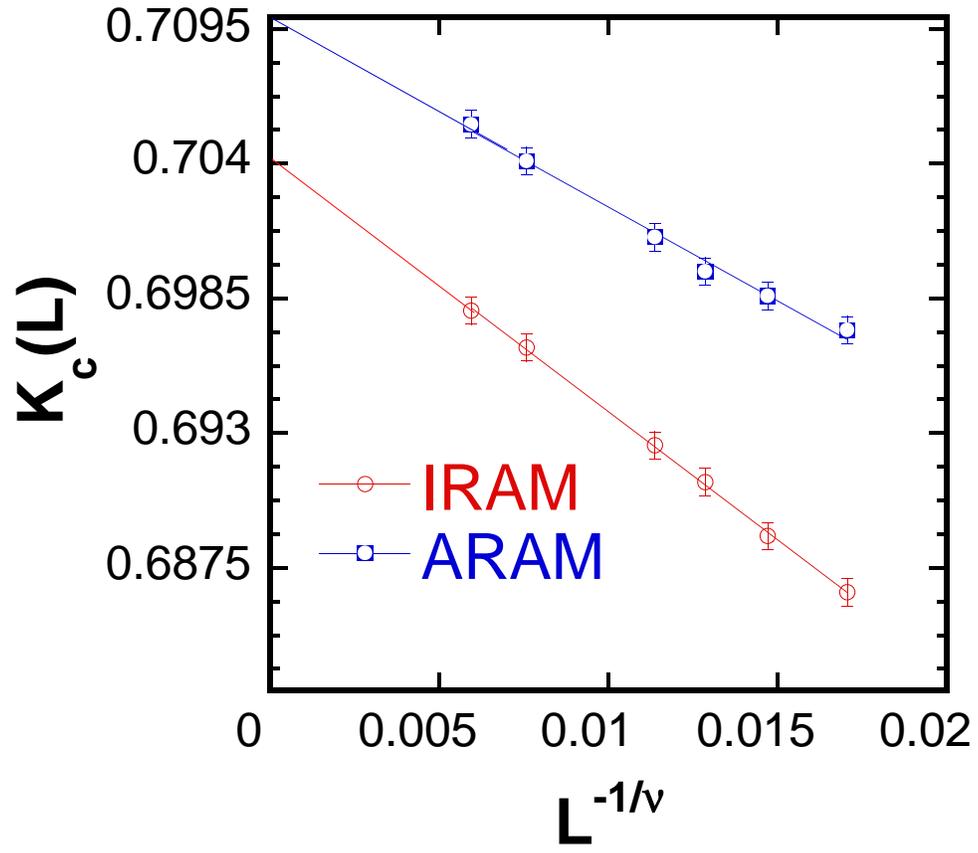}}         
        \end{center}
          \caption{Plots of $K_{c}(L)$ vs. $L^{-1/\nu}$ for IRAM (circle) and ARAM (square).}
       \label{fig.1}  
   \end{figure}

We first aim to determine the critical temperature, $T_{\rm c}$, and the critical exponent of the correlation length, $\nu$. We perform our important-sampling MC simulations with the single histogram method (SHM)\cite{Landaubook,Ferrenberg}. Technically, it has been already optimal to use cluster-flip MC simulations (CFMC) and the multiple histogram method (MHM) to determine $T_{\rm c}$ and $\nu$ independently in the pure Heisenberg model \cite{Landaubook,Ferrenberg}. Unfortunately, neither CFMC nor MHM can be applied efficiently for systems of the RA quenched disorder. The reason for the CFMC case has been pointed out by R\"o\ss{}ler \cite{Robler1999}. For a given couple of lattice sizes, the intersection of their corresponding Binder's cumulant curves varies strongly when changing the set of quenched disorder variables $\{a_{i}\}_{i=1}^{L^3}$. Consequently, it is expensive to carry out several runs with various sets of $\{a_{i}\}_{i=1}^{L^3}$ at different temperatures. Instead, we use the finite size effect (FSE) analysis for the maximum temperature, $T_{\rm c}(L)$, of the susceptibility $\chi$ for each lattice size in the range $18\le L\le40$ to determine simultaneously $T_{\rm c}$ and $\nu$ with moderate efforts. Our simulation detail is as follows. For each lattice size, we run simulations for a modest number of equilibrium configurations (ECs) at different temperatures with the same fixed set of $\{a_{i}\}_{i=1}^{L^3}$ to calculate $\chi(T)$ as a function of the temperature in order to roughly estimate $T_{\rm c}(L)$. We then carry out only a single long-run simulation at the estimated $T_{\rm c}(L)$ and store $10^{7}$ ECs 10 MC sweeps (MCS) apart. These stored ECs are subjected to the single histogram sampling as in \cite{Landaubook,Ferrenberg} to calculate $\chi(T)$ and then to determine precisely $T_{\rm c}(L)$ for each set of $\{a_{i}\}_{i=1}^{L^3}$. Because of the nature of the SHM, we prefer the $K$-representation for the sake of convenience, where $K=J/T$ is called the temperature coupling.   
Fig. \ref{fig.1} presents the size dependence of $K_{\rm c}(L)$.  Each data point in the figure is averaged over ten independent sets of $\{a_{i}\}_{i=1}^{L^3}$. Apparently, the correction-to-finite-size formula $K_{\rm c}(L)=K_{\rm c}+aL^{-1/\nu}$ describes quite well the asymptotic behavior of FSE without the need of the additional correction term $L^{-\omega}$ for both IRAM and ARAM cases. The best fit of this formula to the calculated data yields the values of the critical coupling (temperature) $K_{\rm c}=0.70435(2)$ ($T_{\rm c}=1.41974(4)$) and $K_{\rm c}=0.70998(4)$ ($T_{\rm c}=1.40849(0)$) for the former and the latter cases, respectively. While the value of the critical exponent of the correlation length is the same, i.e $1/\nu=1.40824(0)$ or $\nu=0.70998(0)$, for both cases.

\begin{figure}
       \begin{center}
         \resizebox{!}{100 mm}{\includegraphics{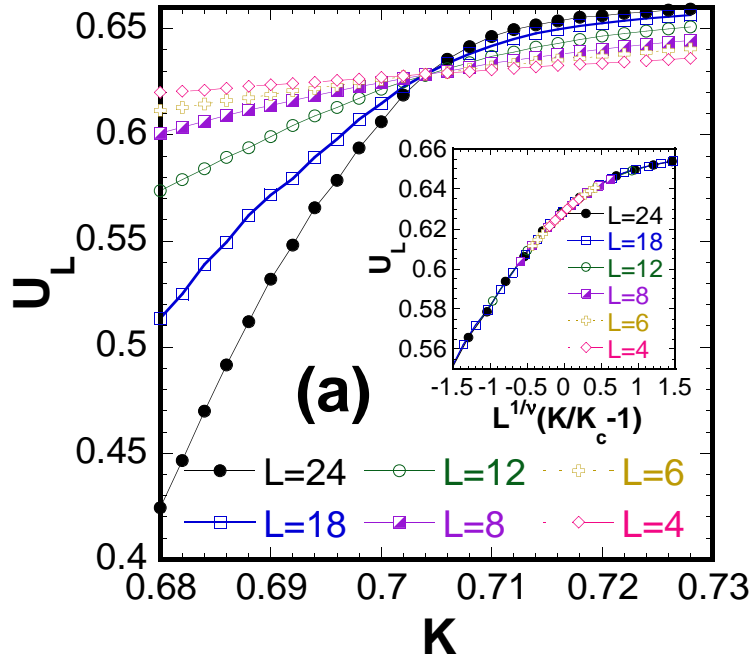}}
         \resizebox{!}{100 mm}{\includegraphics{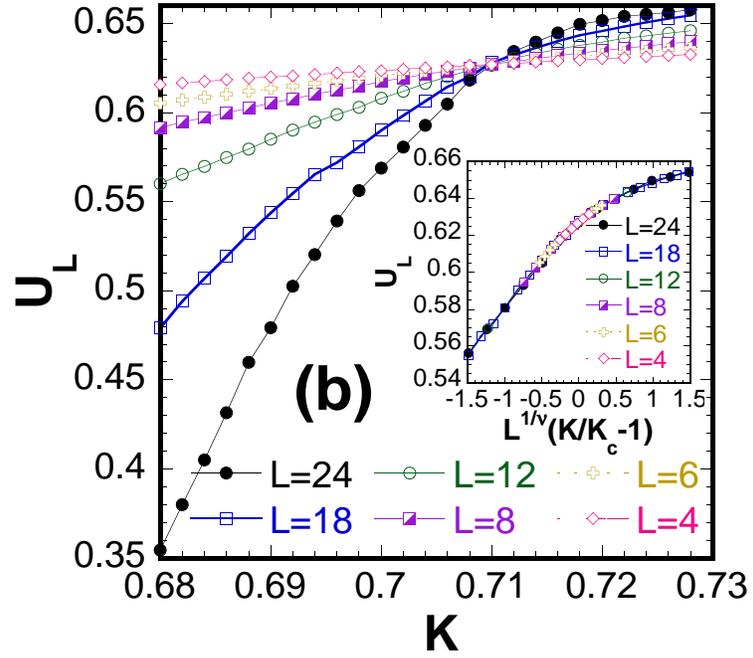}}
        \end{center}
          \caption{Binder's cumulant curves $U_{L}(K)$ for IRAM (a) and ARAM (b). The insets show that the rescaled Binder's cumulant curves overlap into a universal curve for either case. }
       \label{fig.2}  
   \end{figure}

\begin{figure}
       \begin{center}
         \resizebox{!}{!}{\includegraphics{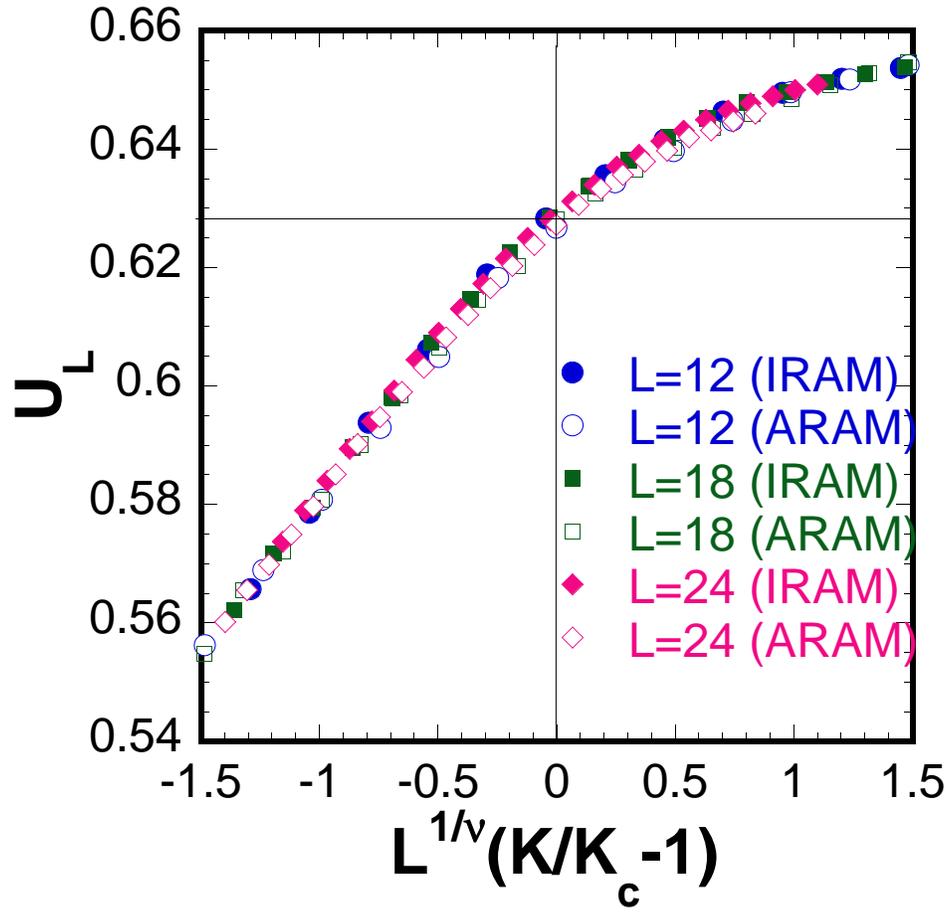}}
        \end{center}
          \caption{The collapse of rescaled Binder's cumulant curves for both IRAM and ARAM. Systems of IRAM and ARM for the case $D/J=4$ may belong to the same universality.}
       \label{fig.3}  
   \end{figure}

Furthermore, we calculate Binder's cumulant $U_{L}(K)=1-\langle m^{4}(K)\rangle/3\langle m^{2}(K)\rangle^{2}$ as a function of $K$ for lattice sizes $4\le L\le 24$ from the simulations of $10^{4}$ statistically independent ECs at various temperature couplings about $K_{\rm c}$. The calculated data, which are averaged over a large number of sets $\{a_{i}\}_{i=1}^{L^3}$, and their rescaled data $\mathcal{U}[L^{1/\nu}(K/K_{\rm c}-1)]$ using the corresponding values of $K_{\rm c}$ and $1/\nu$ above are shown in Fig. \ref{fig.2}. It is remarked from the figure that for either case of the easy-axis distributions there is a well-defined intersection of the family of Binder's cumulant curves which indicates the fixed point of a second-order MPT. The Binder's cumulant, as a thermodynamic function, obeys the same finite-size scaling theory developed by Landau and Binder \cite{Landaubook} as it does in the  pure Heisenberg model. Strikingly, the rescaled Binder's cumulant curves for both IRAM and ARAM collapse nicely into a universal curve shown in Fig. \ref{fig.3}. This fact means that systems of both types of the distributions belong to the same universality class with the correlation length's critical exponent $\nu=0.70998(0)$.

Since the work by Aharony \cite{Aharony} about the `runaway' behavior, i.e. no second-order transition in the RAM in Eq. (\ref{eq1}) for any dimension $d<4$, the nature of phase transition of the model has become a longstanding question. Holovatch {\it et al.} have more recently showed theoretically that the second-order MPT is destroyed by the RA with the distribution in Eq. (\ref{eq2}) but survives in RAM with the distribution in Eq. (\ref{eq3}). However, Monte Carlo simulations by R\"o\ss{}ler \cite{Robler1999}, by Itakura \cite{Itakura2003} and in our present work apparently contrast to the conclusion. In particular, we find nothing different in terms of the critical behavior between IRAM and ARAM as shown above. It is concluded that our present study together with R\"o\ss{}ler's work \cite{Robler1999} shed the light to the possibility of a second-order MPT in the 3D RAMs of weak RA, which is in favor with the existence of the QLRO in 3D weak RAM predicted in the recent FRG theory by Feldman \cite{Feldman2000}. We hope that our work will stimulate further Monte Carlo simulations on issues of the QLRO and the phase transition to it.

This work was financially supported by the National Science Council of Taiwan, R.O.C, under  Grant No. NSC 97-2112-M-007-007-MY3. The computing resources of the National Center for High-Performance Computing under the project ``Taiwan Knowledge Innovation National Grid" are acknowledged.

\end{document}